\documentclass[showpacs,aps,pra,twocolumn]{revtex4}
\usepackage{graphicx,epsfig}
\usepackage{times}
\usepackage{amsmath,amssymb,wasysym}

\newlength\figurewidth
\newcommand{\quoting}[1]{``#1''}
\newcommand{\vc}[1]{\vec{#1}}
\newcommand{\rem}[1]{}

\newcommand{\imagc}[1]{\text{Im}\,#1}

\newcommand{\realb}[1]{\text{Re}[#1]}
\newcommand{\realc}[1]{\text{Re}\,#1}

\begin{document}

\title{Asymmetric scattering and non-orthogonal mode patterns in optical micro-spirals}
  \author{Jan Wiersig}
  \affiliation{Institut f{\"u}r Theoretische Physik, Universit{\"a}t Magdeburg,
  Postfach 4120, D-39016 Magdeburg, Germany}
  \author{Sang Wook Kim}
\affiliation{Departement of Physics Education, Pusan National University, Busan 609-735, Korea}
  \author{Martina Hentschel}
\affiliation{Max-Planck-Institut f\"ur Physik komplexer Systeme, N{\"o}thnitzer
Stra{\ss }e 38, D-01187
Dresden, Germany}
\date{\today}
\begin{abstract}
Quasi-bound states in an open system do in general not form an orthogonal and
complete basis. It is, however, expected that the non-orthogonality is weak
in the case of well-confined states except close to a so-called exceptional
point in parameter space. We present numerical evidence showing that for
passive optical  microspiral cavities the parameter regime where the
non-orthogonality is significant is rather broad. Here we observe
almost-degenerate pairs of well-confined modes which are highly non-orthogonal.
Using a non-Hermitian model Hamiltonian we demonstrate that this interesting
phenomenon is related to the asymmetric scattering between clockwise and
counterclockwise propagating waves in the spiral geometry. Numerical
simulations of ray dynamics reveal a clear ray-wave correspondence.
\end{abstract}
\maketitle

\section{Introduction}
Dielectric microdisks and spheres have been extensively studied due to the
extraordinarily high quality factors ($Q$-factors) achieved in such
structures~\cite{Vahala03}. Such a
merit is ascribed to the formation of so-called whispering gallery (WG) modes 
based on total internal reflection upon the surfaces of circular or
spherical cavities~\cite{Chang96}. However, these highly symmetrical shapes
prohibit a directional light output, which is clearly
disadvantageous for practical applications. An obvious solution is to break such symmetries by deforming the cavity shape~\cite{Nockel97,Gmachl98,Stone03}. The final goal in this direction is to obtain {\em unidirectional} output without degrading the high $Q$-factor too much~\cite{Chern03,Kurdoglyan04,Kneissl04,Wiersig06,Wiersig08}.

\begin{figure}
\includegraphics[width=0.6\figurewidth]{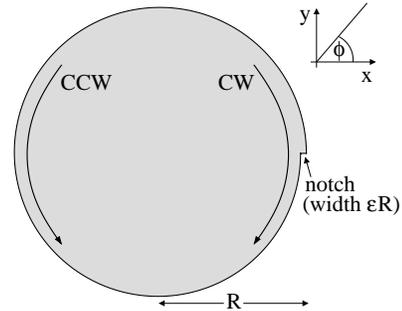}
\caption{Schematic top view of the spiral-shaped cavity. Clockwise (CW) and counterclockwise (CCW) traveling waves are indicated.}
\label{fig:sketch}
\end{figure}

Chern {\it et al.} experimentally demonstrated unidirectional lasing from a
spiral-shaped microcavity~\cite{Chern03}, see the sketch in
Fig.~\ref{fig:sketch}. The maximum emission comes from the notch of the spiral
at an angle $\phi$ of about $-35^{\circ}$~\cite{Chern03,Kneissl04,HKBC08} or
some different angle~\cite{Messaoud05,Fujii05,Fujii06,Audet07}. 
The possible origins of these different results are discussed in Ref~\cite{HKBC08}.
It is clear that one characteristic feature of the spiral cavity plays a
crucial role for the emission directionality: the chiral
symmetry is broken so that the clockwise (CW) rotation is distinct from the
counter-clockwise (CCW) rotation~\cite{Chern03}. Recently, the appearance of
pairs of almost-degenerate modes with mainly CCW character but also small CW
component has been observed~\cite{Wiersig08oe}.
Another characteristic feature of the spiral-shaped cavity is the existence of
so-called quasi-scarred modes~\cite{Lee04,Kwon06}. These interesting modes
seem to be localized along simple periodic ray trajectories. A careful
analysis, however, shows that although these simple periodic trajectories do
not exist in the conventional ray dynamics, they do exist in an augmented ray 
dynamics including Fresnel filtering~\cite{EMH08}.

Microcavities are open wave systems which are described not by Hermitian but 
by non-Hermitian operators~\cite{Okolowicz03}, whose distinct properties are 
that their eigenvalues are in general complex and the eigenvectors are not
orthogonal. The imaginary parts of the eigenvalues have the physical meaning
of the decay rates of the modes. The modes of the spiral cavity can (in
principle) be obtained by diagonalizing a non-Hermitian matrix so that they
mutually form a non-orthogonal set.  
Another pronounced feature of the non-Hermitian matrix is the existence of 
exceptional points (EPs) in parameter space. At an EP (at least) two 
eigenvalues coalesce~\cite{Kato66,Heiss00} and the corresponding two 
eigenvectors converge to each other so that the dimension of the eigenbasis of
the non-Hermitian matrix is reduced~\cite{Heiss04}. 
It is known that not only does the non-trivial topology of complex eigenvalues
around the EP exist~\cite{Dembowski01} but also a relation to Berry's geometric
phase which appears when external parameters are varied 
adiabatically~\cite{Dembowski03,Dembowski04,Mailybaev05}.  Recently the
non-Hermitian matrix and the EP have attracted much
interest~\cite{Wiersig06b,Unterhinninghofen08,Dietz07,Cartarius07,Keating08}. 

In this paper, we draw a connection from the appearance of almost-degenerate modes with
mainly CCW character~\cite{Wiersig08oe} to the broken chiral symmetry
by the asymmetric scattering between CCW and CW traveling waves. We
show that an effective 2-by-2 non-Hermitian matrix is extremely useful to
understand the main characteristics of this relationship. When the effective
transition rate from the CCW to the CW component, i.e. an off-diagonal element of the
non-Hermitian matrix, vanishes, two eigenvalues of the matrix
coalesce to form the EP, at which there exists only one eigenvector. The
physical meaning of this is transparently revealed in the spiral cavities in
the context of the time evolution of the scattering dynamics.

In Sec. II, we introduce the spiral cavity. Section III reports our numerical results on the properties of optical modes in this kind of cavity. In Sec. IV we introduce an effective non-Hermitian Hamiltonian to describe the non-orthogonality. The time evolution of waves in discussed in Sec. V. The ray-wave correspondence is the subject of Sec. VI. In Sec. VII we discuss the dependence of the mode properties on a shape parameter. A summary is given in Sec. VIII.

\section{The system}

Microdisk cavities are quasi-two-dimensional systems with
piece-wise constant effective index of refraction $n(x,y)$. In this case
Maxwell's equations reduce to a two-dimensional scalar mode
equation~\cite{Jackson83eng}
\begin{equation}\label{eq:wave} -\nabla^2\psi =
n^2(x,y)\frac{\omega^2}{c^2}\psi \ ,
\end{equation}
with frequency $\omega = ck$, wave number
$k$, and the speed of light in vacuum $c$. The mode equation~(\ref{eq:wave})
is valid for both transverse magnetic (TM) and transverse electric (TE)
polarization. For TM polarization the electric field $\vec{E}(x,y,t) \propto
(0,0,\realb{\psi(x,y)e^{-i\omega t}})$ is perpendicular to the cavity plane.
The wave function $\psi$ and its normal derivative
are continuous across the boundary of the cavity. For TE
polarization, $\psi$ represents the $z$-component of the magnetic field vector
$H_z$. Again, the wave function $\psi$ is continuous across the boundaries,
but its normal derivative $\partial_\nu\psi$ is not. Instead,
$n(x,y)^{-2}\partial_\nu\psi$ is continuous~\cite{Jackson83eng}.
At infinity, outgoing wave conditions are imposed which results in quasi-bound
states with complex
frequencies $\omega$ in the lower half-plane. Whereas the real part is the
usual frequency, the imaginary part is related to the lifetime
$\tau=-1/[2\,\imagc{\omega}]$ and to the quality factor
$Q = -\realc{\omega}/[2\,\imagc{\omega}]$.

If the domain of interest is simply connected then the mode equation~(\ref{eq:wave}) can be written as the following
eigenvalue equation
\begin{equation}\label{eq:eigenvalue0}
\left(-\frac{c^2}{n^2}\nabla^2\right)\psi = \omega^2\psi \ .
\end{equation}
In a closed cavity with vanishing wave function along the boundary the linear operator
on the left hand side is Hermitian. In the case of optical microcavities, however, the operator is non-Hermitian
because of the outgoing wave conditions.

In polar coordinates $(r,\phi)$ the boundary of the spiral cavity is defined as
\begin{equation}
r(\phi) = R\left(1-\frac{\varepsilon}{2\pi}\phi\right)
\end{equation}
 with deformation parameter $\varepsilon \geq 0$ and ``radius'' $R>0$ at
$\phi = 0$ as shown in Fig.~\ref{fig:sketch}.
The radius jumps back to $R$ at $\phi = 2\pi$ creating a notch. Note that $R$ 
is a trivial parameter which can be scaled away by using
normalized frequencies $\Omega = \omega R/c = kR$. We are left with a one-parameter family of boundary shapes parametrized by $\varepsilon$.
As in Ref.~\cite{Wiersig08oe} we consider TE polarization, an effective index
of refraction $n=2$ for silicon nitride, and, if not otherwise stated, the relative notch width $\varepsilon = 0.04$.

\section{Mode properties}
\label{sec:numresults} We use the boundary element method~\cite{Wiersig02b} to compute the spatial mode patterns $\psi(x,y)$ and the complex frequencies $\Omega$. Figure~\ref{fig:mode1} shows an interesting example, a pair of {\it nearly} degenerate modes, which are of quasi-scar type~\cite{Lee04,Kwon06}. One of these modes has $\Omega = 41.4676-i0.03422$, i.e. a quality factor $Q_1 = 606$ and, if we assume  $R = 10\,\mu$m, a free-space wavelength $\lambda_1 = 1515.2\,$nm. The other mode has $\Omega = 41.4627-i0.03473$ corresponding to $\lambda_2 =
1515.38\,$nm and $Q_2 = 597$. Not only the resonant wavelength and $Q$-factor
are very similar, but also the mode patterns depicted in Fig.~\ref{fig:mode1}. The question arises whether these numerical solutions really correspond to different modes. That this is indeed the case can be seen for instance in the far-field intensity patterns in Fig.~{\ref{fig:farfield1}. We can observe small oscillations with a phase difference of $\pi$ superimposed on the common envelope.
\begin{figure}
\includegraphics[width=0.95\figurewidth]{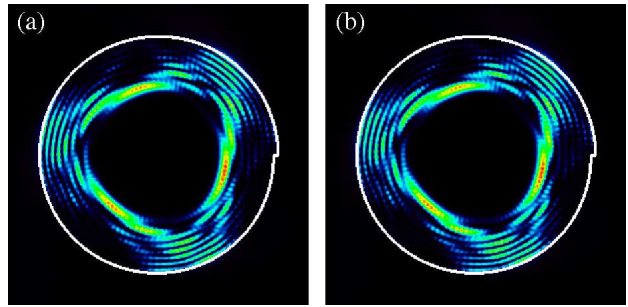}
\caption{(Color online) Calculated intensity $|\psi|^2$ of the nearly degenerate quasi-scar modes 1 (a) and 2
  (b).}
\label{fig:mode1}
\end{figure}
\begin{figure}
\includegraphics[width=0.95\figurewidth]{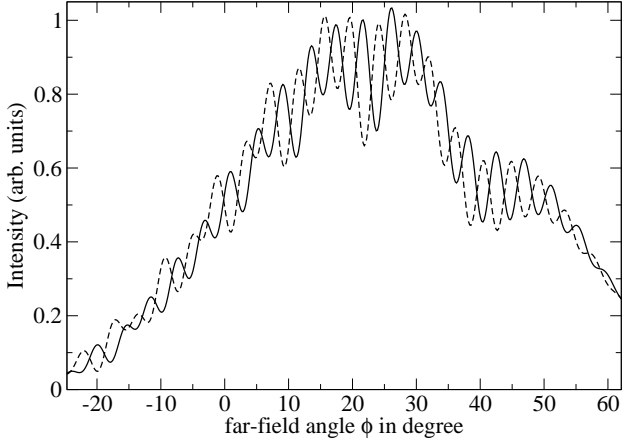}
\caption{Magnification of the computed far-field patterns of the modes in
  Fig.~\ref{fig:mode1}. The solid (dashed) line corresponds to mode 1 (2). For the definition of the far-field angle $\phi$ see Fig.~\ref{fig:sketch}.} \label{fig:farfield1}
\end{figure}

Following Refs.~\cite{Wiersig08oe,Chern03} we analyze the mode pattern by
expanding the wave function inside the cavity in cylindrical harmonics
\begin{equation}
\psi(r,\phi) = \sum_{m=-\infty}^{\infty} \alpha_m J_m(nkr)\exp{(im\phi)}
\end{equation}
where $J_m$ is the $m$th order Bessel function of the first kind. Positive (negative) values of the angular
momentum index $m$ correspond to CCW (CW) traveling-wave components. In
Fig.~\ref{fig:AMD}(a) we can observe that for both modes the angular momentum
distribution $|\alpha_m|^2$ is dominated by the CCW component, i.e. none of
the two modes can be classified as CW traveling-wave mode. The tiny difference
between the modes can be seen in Figs.~\ref{fig:AMD}(b) and (c). For negative
angular momentum index both the real and the imaginary part of $\alpha_m$ have
a different sign for the two modes which is nothing else than the phase
difference of $\pi$ observed in the far-field pattern. That means, we can
construct superpositions with $\alpha^\pm_m = (\alpha^{(1)}_m\pm
\alpha^{(2)}_m)/2$ being CW and CCW traveling-waves, respectively, as can be
seen in Fig.~\ref{fig:AMD}(d). Experimentally, the selective excitation of such traveling waves can be done by coupling light into the cavity via an attached waveguide~\cite{LLP07}. It is important to emphasize that these superpositions are not eigenmodes of the cavity as they are composed of two modes with slightly different frequencies and $Q$-factors.

\begin{figure}
\includegraphics[width=1.0\figurewidth]{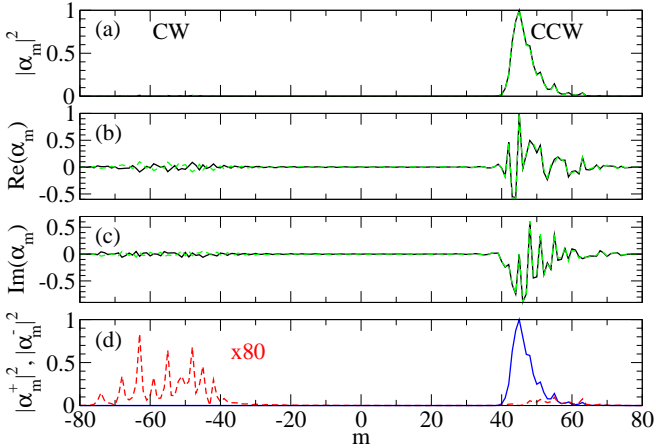}
\caption{(Color online) Angular momentum distributions $\alpha^{(1)}_m$ (solid line) and $\alpha^{(2)}_m$ (dashed) of quasi-scar modes (see Fig.~\ref{fig:mode1}) normalized to 1 at maximum: (a) absolute
value squared, (b) real and (c) imaginary part. (d) Superpositions $\alpha^+_m = (\alpha^{(1)}_m+ \alpha^{(2)}_m)/2$ (solid) and $\alpha^-_m
= (\alpha^{(1)}_m-\alpha^{(2)}_m)/2$ (dashed, multiplied by a factor of 80).}
\label{fig:AMD}
\end{figure}

The discussed properties of the angular momentum distributions are not
restricted to quasi-scar modes. Figure~\ref{fig:mode2} depicts an example of a
pair of \quoting{WG-like} modes in the same cavity. One mode has $\Omega =
41.7166-i0.00207$, i.e. a quality factor $Q_1 =  10067$ and, if we again
assume  $R = 10\,\mu$m, a free-space wavelength $\lambda_1 = 1506.16\,$nm. The
other mode has $\Omega = 41.7193-i0.00184$ corresponding to $\lambda_2 =
1506.06\,$nm and $Q_2 = 11363$. Inspection of the angular momentum
distributions in Fig.~\ref{fig:Y1} shows qualitatively the same scenario as
for the quasi-scars, even though the asymmetry between CW and CCW components
is much weaker.
\begin{figure}
\includegraphics[width=0.95\figurewidth]{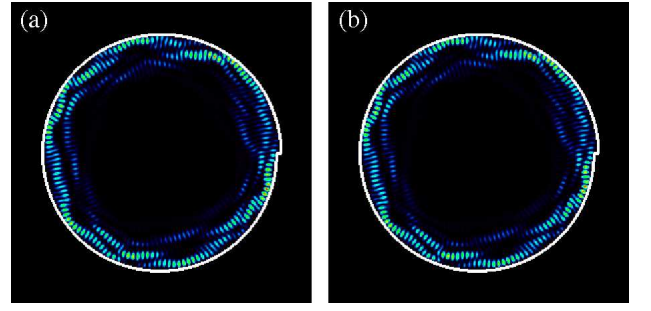}
\caption{(Color online) Calculated intensity $|\psi|^2$ of the nearly degenerate WG-like modes 1 (a) and 2
  (b).}
\label{fig:mode2}
\end{figure}
\begin{figure}
\includegraphics[width=1.0\figurewidth]{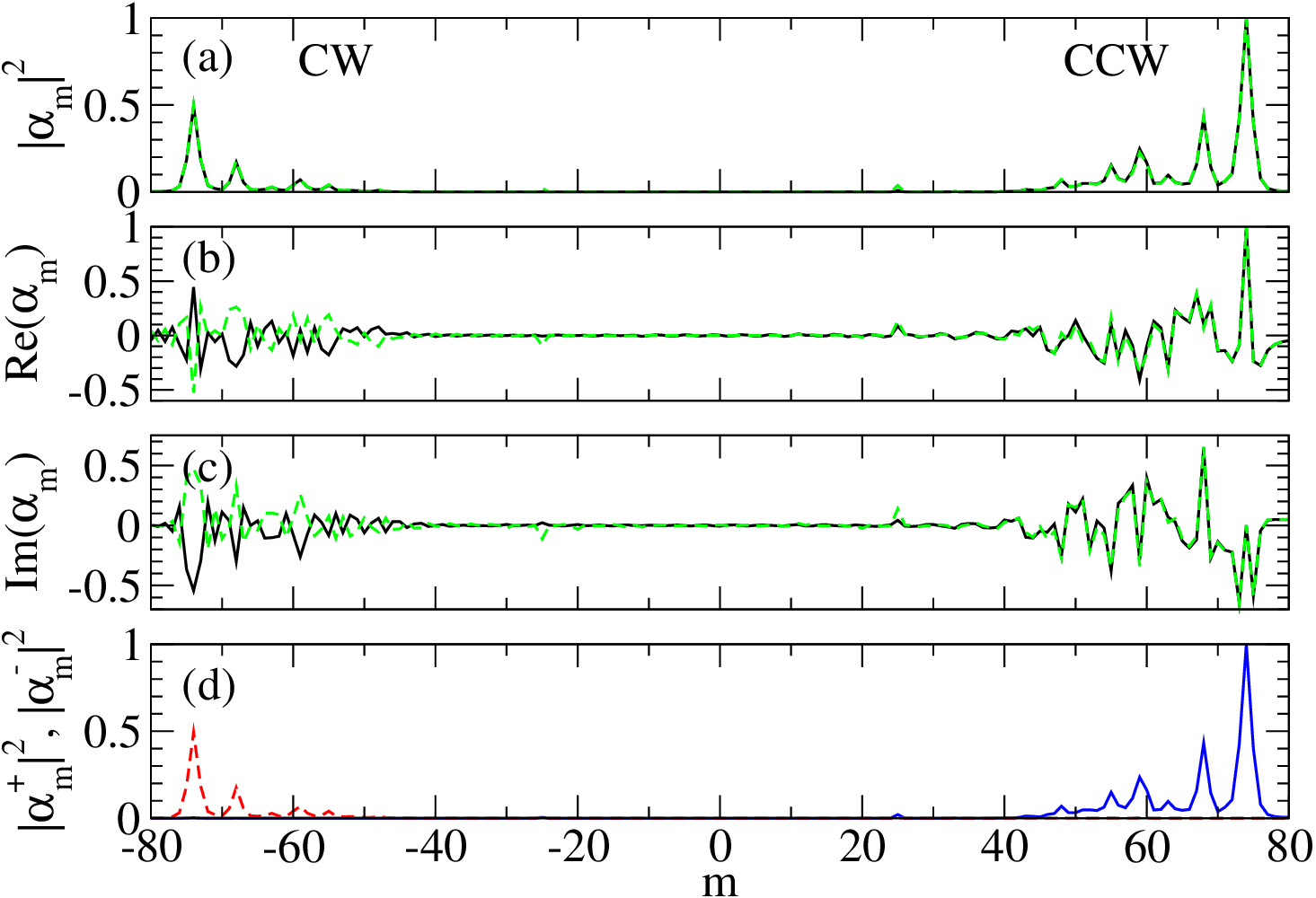}
\caption{(Color online) Angular momentum distributions $\alpha^{(1)}_m$ (solid line) and $\alpha^{(2)}_m$ (dashed) of WG-likes modes (see Fig.~\ref{fig:mode2}) normalized to 1 at maximum:
(a) absolute value squared, (b) real and (c) imaginary part. (d) Superpositions $\alpha^+_m = (\alpha^{(1)}_m+ \alpha^{(2)}_m)/2$ (solid) and
$\alpha^-_m = (\alpha^{(1)}_m-\alpha^{(2)}_m)/2$ (dashed).} \label{fig:Y1}
\end{figure}

This kind of almost-degenerate modes in spiral-shaped cavities have already
been observed in Ref.~\cite{Wiersig08oe}. Here, we will discuss new features such as non-orthogonality of these modes and the implication on the time evolution of CW and CCW traveling waves in such cavities. To quantify the
non-orthogonality we compute
the normalized overlap integral of two modes $\psi_1$ and $\psi_2$ over the interior of the cavity~${\cal C}$
\begin{equation}
\label{eq:overlap}
S = \frac{|\int_{\cal C} dxdy\;\psi_1^*\psi_2|}{\sqrt{\int_{\cal C} dxdy\;\psi_1^*\psi_1}\sqrt{\int_{\cal C} dxdy\;\psi_2^*\psi_2}} \ .
\end{equation}
For the quasi-scar modes in Fig.~\ref{fig:mode1} we find $S\approx 0.972$. This value close to unity means that the modes are nearly collinear.
The overlap for the WG-like modes in Fig.~\ref{fig:mode2} is $S\approx 0.481$, i.e. the non-orthogonality is weaker but still quite significant.
Note that the overlap of modes from different pairs, i.e. one mode from
Fig.~\ref{fig:mode1} and one from Fig.~\ref{fig:mode2}, is numerically around 1
percent or below. That means, the modes are only pairwise non-orthogonal.

We have studied in total 60 modes in this cavity geometry within various frequency regimes and also for TM polarization. We always find that the modes come as strongly non-orthogonal pairs. We therefore conjecture that this is a generic property of optical modes in the spiral microcavity.

\section{Effective non-Hermitian Hamiltonian}
In this section we demonstrate that the discussed behavior of the pairs of modes can
be modeled by a simple toy model, the 2-by-2 non-Hermitian and non-symmetric
matrix
\begin{equation}\label{eq:nonHermitianMatrix}
H = \left(\begin{array}{cc} E_0 & 0 \\ 0 & E_0 \end{array}\right) +
    \left(\begin{array}{cc} \Gamma & V \\ \eta V^* & \Gamma \end{array}\right) \ .
\end{equation}
The eigenvectors of the first matrix on the r.h.s. belong to the CCW and CW
traveling waves with equal wave number $\sqrt{E_0}\in\mathbb{C}$ in the
absence of any coupling between them. The first matrix suffices to describe the situation for the circular cavity ($\varepsilon=0$). In the general case of $\varepsilon>0$ the second matrix accounts for coupling effects which can be interpreted as scattering between CCW and CW traveling waves. The diagonal elements are given by the total scattering rates $\Gamma$ which are assumed to be equal for simplicity. The off-diagonal element $V = |V|e^{i\delta} \in\mathbb{C}$ describes scattering from a CW traveling wave to the CCW traveling wave. The other off-diagonal element $\eta V^*$ describes scattering from a CCW traveling wave to the CW traveling wave. The latter scattering is assumed to be weaker, i.e. $0\leq\eta<1$, which is reasonable because of the geometry of the system, see Fig.~\ref{fig:sketch}. 
The complex eigenvalues of the matrix~(\ref{eq:nonHermitianMatrix}) are given by
\begin{equation}
E_\pm = E_0+\Gamma\pm\sqrt{\eta}|V| \ . \label{eq:eigenvalue}
\end{equation}
The right eigenvectors turn out to be
\begin{equation}
\vec{\alpha}_\pm = \frac{1}{\sqrt{2}}\left(\begin{array}{c} 1 \\ \pm\sqrt{\eta}e^{-i\delta}\end{array}\right) \ . \label{eq:eigenvector}
\end{equation}
These eigenvectors directly explain the mode structure discussed in
Section~\ref{sec:numresults}. The weight of the first component (corresponding
to CCW traveling waves) $\sim 1$ is much larger than that of the second 
component
(corresponding to CW traveling waves)  $\sim \eta$, cf. Figs.~\ref{fig:AMD}
and \ref{fig:Y1}.

The two eigenvectors in Eq.~(\ref{eq:eigenvector}) are non-orthogonal in the
case of asymmetric scattering ($\eta\neq 1$):
\begin{equation}\label{eq:nonortho}
\frac{|\vec{\alpha}_+^*\cdot\vec{\alpha}_-|}{|\vec{\alpha}_+||\vec{\alpha}_-|} = \frac{1-\eta}{1+\eta} \ .
\end{equation}
Applying Eq.~(\ref{eq:nonortho}) to the pair of quasi-scars in
Fig.~\ref{fig:mode1} with $S\approx 0.972$ yields $\eta^{-1} \approx 71$. This 
high degree of asymmetry in the scattering is consistent with the scaling
factor of $80$ in Fig.~\ref{fig:AMD} using the connection to the mode
structure in Eq.~(\ref{eq:eigenvector}). In the case of the WG-like mode in
Fig.~\ref{fig:mode2} with $S\approx 0.481$ we get $\eta^{-1} \approx
2.8$. This lower degree of asymmetry in the scattering is in agreement with the fact that the CW component is only by a 
factor of about 2 smaller than the CCW component in Fig.~\ref{fig:Y1}.

The left eigenvectors of the matrix~(\ref{eq:nonHermitianMatrix}) are given by
\begin{equation}
\vec{\beta}_\pm =  \frac{1}{\sqrt{2}}\left(1, \pm\frac{1}{\sqrt{\eta}}e^{i\delta} \right) \ .
\end{equation}
Left and right eigenvectors are orthogonal to each other, i.e. $\vec{\beta}_\pm\cdot\vec{\alpha}_\mp = 0$ and
$\vec{\beta}_\pm\cdot\vec{\alpha}_\pm = 1$, as can be easily verified. Note that $\vec{\beta}_\pm^*\neq\vec{\alpha}_\pm$.

A remark is in order. In an open resonator the laser linewidth is increased
with respect to the well-known Shawlow-Townes formula by the so-called
Petermann factor~$K$~\cite{Petermann79,FSPB00,SFPB00}. At an EP the Petermann factor is expected to diverge~\cite{LRS08}.
For the discussed non-Hermitian Hamiltonian the Petermann factor $K$ is
obtained as
\begin{equation}
K=\frac{\left<\beta_\pm|\beta_\pm\right> \left<\alpha_\pm|\alpha_\pm\right>}{\left| \left<
\beta_\pm|\alpha_\pm\right>\right|^2}=\frac{1}{4}\left(1+\frac{1}{\eta}\right)(1+\eta),
\end{equation}
which diverges as $\eta \rightarrow 0$. This property originates from the
non-orthogonality of eigenstates of the non-Hermitian matrix. Interestingly,
the Petermann factor~$K$ here depends solely on the degree of asymmetry~$\eta$ 
of the scattering between CW and CCW traveling waves.

\section{Time evolution of CW and CCW traveling waves}
The time evolution of waves is determined by the time-dependent Maxwell's
equations which for a quasi-2D system can be written as
\begin{equation}\label{eq:Maxwell}
\ddot{\Psi} = -H\Psi.
\end{equation}
Being of second-order this two-component differential equation has four
independent solutions proportional to the eigenvectors of $H$. Two solutions
$\vc{\alpha}_\pm e^{-i\sqrt{E_\pm}t}$ with positive frequency,
$\realc{\sqrt{E_\pm}} > 0$, and two of the same kind with negative frequency.
The solutions with negative
frequency are not considered here separately as they come naturally into
play when the real part of the wave function is taken to determine the
$z$-component of the magnetic field as
\begin{eqnarray}
H_z(x,y,t) & \propto & \realb{\psi(x,y)e^{-i\omega t}}\\
& \propto & \frac{1}{2}\left[\psi(x,y)e^{-i\omega t}+\psi(x,y)^*e^{i\omega t}\right] \ .
\end{eqnarray}

From Eq.~(\ref{eq:eigenvector}) it is clear that at
$\eta = 0$ we have an EP: the two right eigenvectors $\vec{\alpha}_\pm$
collapse into a single vector $(1,0)^T$, where the superscript $T$ represents the transpose of the matrix. This eigenvector
represents a pure CCW traveling wave. One can then ask the following questions: What is its physical implication? What happens if the initial state is
chosen as $(0,1)^T$, i.e. normal to the eigenvector? To address these questions we
rewrite the differential equation~(\ref{eq:Maxwell}) in the following form by using $\Psi = (a,b)^T$.
\begin{eqnarray}
\ddot{a} & = & -Ea - Vb, \\ %
\ddot{b} & = & - Eb, %
\label{eq:ep}
\end{eqnarray}
where $E=E_0+\Gamma$.  One solution is obviously
\begin{equation}
\Psi_1 = \left(
\begin{array}{c}
e^{-i\sqrt{E}t} \\
0
\end{array}
\right), \label{eq:Psi1}
\end{equation}
which exhibits an exponential decay in time, i.e. $|\Psi_1| \sim
e^{-\gamma t}$, where $\gamma =-\imagc{\sqrt{E}} > 0$ and $\realc{\sqrt{E}} >
0$.
Note that the solution in Eq.~(\ref{eq:Psi1}) is proportional to the
eigenvector $(1,0)^T$ of the Hamiltonian.
One then expects $(0,1)^T$ could be another eigenvector. However, it is not true. When $(0,1)^T$ is chosen as the initial condition, one obtains
\begin{equation}
\ddot{a} + Ea = -Ve^{-i\sqrt{E}t},%
\label{eq:01}
\end{equation}
where the r.h.s. comes from Eq.~(\ref{eq:ep}), which is independent of $a$. It is easy to see that
\begin{equation}
a(t) = \frac{V}{2i\sqrt{E}}te^{-i\sqrt{E}t}
\end{equation}
 is a solution of Eq.~(\ref{eq:01}). Finally, one obtains
\begin{equation}
\Psi_2 = \left(
\begin{array}{c}
Vt/(2i\sqrt{E})\\%
1
\end{array}
\right)e^{-i\sqrt{E}t}. \label{eq:Psi2}
\end{equation}
Note that $\Psi_2$ is not an eigenvector of $H$ nor a simple harmonic function
of time, but satisfies $(0,1)^T$ at
$t=0$. Again there is an analogue solution with negative frequency which is
not considered here. One conclusion drawn from our analysis is that the
differential equation~(\ref{eq:Maxwell}) has two independent solutions (plus
two with negative frequency) also at the EP, even though the eigenspace of the
Hamiltonian $H$ has only one dimension in this case.

The physical meaning of the two solutions (\ref{eq:Psi1}) and
(\ref{eq:Psi2}) is clear in the spiral cavity. For $\Psi_1$ the state is initially prepared in the CCW traveling wave, so that it cannot make
any transition to the CW since the corresponding transition rate is zero in the $\eta = 0$ case. This is also intuitively acceptable because the CCW traveling wave
experiences much less scattering than the CW traveling wave; see Fig.~\ref{fig:sketch}. For $\Psi_2$ the state is initially prepared in the CW traveling wave, and it exhibits the
transition to the CCW leading to non-exponential time dependence. The scattering upon the notch gives rise to the
transition from the CW to the CCW traveling wave so that the initial linear increase of the amplitude, see Eq.~(\ref{eq:Psi2}), comes into play. Such a non-exponential
decay has been proposed in different context~\cite{Cardamone02} and
experimentally observed in a microwave cavity~\cite{Dietz07}. Slightly
away from the EP with $\eta \apprge 0$ the two eigenvectors still have
considerable overlap. One can therefore expect a similar behavior to 
occur also in the vicinity of the EP.

To confirm the discussed scattering scenario, we consider superpositions of
modes which can be interpreted as pure CW or CCW traveling waves. As example
we take the quasi-scar modes presented in Fig.~\ref{fig:mode1}. The modes evolve according to their complex frequencies $\omega_j = ck_j$ computed with the boundary element method. The first kind of superposition we consider is given by
\begin{equation}
 \left(\begin{array}{c}\chi^+ \\ \chi^-\end{array}\right) = 
\frac12\left(\begin{array}{c} \alpha_1 \\ \alpha_2 \end{array}\right) e^{-i\omega_1 t} + 
\frac12\left(\begin{array}{c} \alpha_1 \\ -\alpha_2 \end{array}\right) e^{-i\omega_2
  t}
\ .
\end{equation}
For the coefficients $\alpha_j$ we choose as a rough approximation the maximal coefficients of the angular momentum distribution in Figs.~\ref{fig:AMD}(b) and (c): $\alpha_1 = 1$ and $\alpha_2 = 0.1+i0.07$. For $t=0$ the superposition is purely CCW traveling as $\chi^+ = \alpha_1$ and $\chi^- = 0$.  The upper panel of
Fig.~\ref{fig:dynamics} shows the evolution of the superposition. We observe only a weak scattering into the CW component as
it is expected from the 2-by-2 matrix~(\ref{eq:nonHermitianMatrix}) with small $\eta$.
The second kind of superposition is
\begin{equation}
 \left(\begin{array}{c}\chi^+ \\ \chi^-\end{array}\right) = 
\frac12\left(\begin{array}{c} \alpha_1 \\ \alpha_2 \end{array}\right) e^{-i\omega_1 t} + 
\frac12\left(\begin{array}{c} -\alpha_1 \\ \alpha_2 \end{array}\right) e^{-i\omega_2
  t}
\ .
\end{equation}
Initially, this superposition is a pure CW traveling wave since $\chi^+ = 0$ and $\chi^- = \alpha_2$. The lower  panel of Fig.~\ref{fig:dynamics} shows the
evolution of this superposition. Here we observe a strong
scattering into CCW components with a linear increase of the amplitude as
predicted by Eq.~(\ref{eq:Psi2}). In conclusion, the full mode calculation near
the EP demonstrates very similar dynamical behavior as the effective non-Hermitian
Hamiltonian at the EP. This clearly shows that at the EP nothing dramatic
happens to the dynamics, even though the eigenspace of $H$
collapses.
\begin{figure}
\includegraphics[width=1.0\figurewidth]{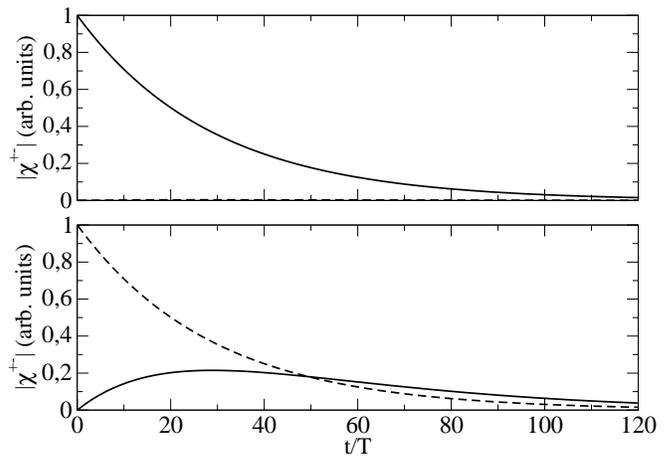}
\caption{Time evolution of CCW (solid lines) and CW (dashed) traveling waves. The traveling waves are superpositions of the quasi-scar modes with frequencies $\Omega_1$ and $\Omega_2$ depicted in Fig.~\ref{fig:mode1}. The upper (lower) panel contains the dynamics starting with a pure
CCW (CW) traveling wave. Time is measured in units of $T = 2\pi/\realc{\Omega_1}$.} \label{fig:dynamics}
\end{figure}

\section{Ray-wave correspondence}
Our previous considerations have revealed that the peculiar time evolution of
CCW and CW traveling waves plotted in Fig.~\ref{fig:dynamics} is related to the
non-orthogonality associated with an EP. As the EP is an intrinsic wave
phenomenon one would not expect to find a ray-wave correspondence in such a
situation.
Our numerical ray simulations, however, prove that this intuitive expectation
fails. The upper panel of Fig.~\ref{fig:rays} has been computed by starting
$10\,000$ rays uniformly at the boundary of the cavity with initial directions
corresponding to initially CCW propagation in the regime of total internal
reflection. During the time evolution the rays can partially leave the cavity 
by refraction whenever the angle of incidence becomes smaller
than the critical angle for total internal reflection. The transmission is
given by Fresnel's laws. Moreover, rays can change their sense of
rotation to CW propagation whenever they hit the notch. However, hitting the
notch is very unlikely for CCW propagation rays as it is intuitively clear from
Fig.~\ref{fig:sketch}. Therefore, practically all rays have completely left
the cavity before they can change the sense of rotation. This is very
different when the rays start initially in the CW sense of propagation as
can be seen in the lower panel of Fig.~\ref{fig:rays}. Here, rays can easily hit the
notch and thereby change their sense of rotation.
Comparison of the ray dynamics in Fig.~\ref{fig:rays} and the wave dynamics
in Fig.~\ref{fig:dynamics} uncovers a striking correspondence. We believe that
this unexpected ray-wave correspondence has two reasons: (i) In both cases the
time evolution is governed by the asymmetric scattering between CW and CCW
traveling components, and (ii) at the EP nothing special happens to the wave
dynamics, despite the collapse of the eigenspace of $H$.
\begin{figure}
\includegraphics[width=1.0\figurewidth]{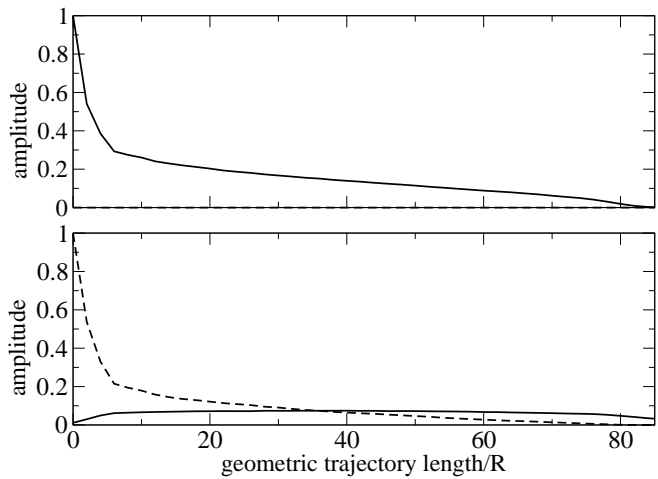}
\caption{Time evolution of amplitude (defined as square root of intensity in 
arbitrary units) corresponding to CCW 
(solid lines) and CW (dashed) propagating light
  rays. The upper (lower) panel shows the dynamics starting with a set of pure
  CCW (CW) propagating rays in full analogy to the wave dynamical
  considerations in Fig.~\ref{fig:dynamics}. Time is proportional to the
  geometric length of ray trajectories.} \label{fig:rays}
\end{figure}


There are several facts which could explain the quantitative differences
between the wave and ray dynamics: (i) The wavelength in the material
$\lambda/n$ is about $0.075R$ which is larger than the notch width
$\varepsilon R = 0.04R$. This is illustrated in Fig.~\ref{fig:magnification}.
(ii) Diffraction at the corners is not included in the ray dynamics. We expect 
that for the WG-like modes diffraction at the corners can be important as the field intensity is larger at the corner than at the notch itself; see Fig.~\ref{fig:magnification}(b). Since diffraction at the corner is more uniform than direct scattering at the notch, see Ref.~\cite{CBW05}, diffraction explains why WG-like modes show less asymmetry (non-orthogonality) if compared to the quasi-scar modes.
(iii) The considered pair of modes are not uniformly
distributed in phase space. (iv) For the coefficients $\alpha_j$ we have
chosen only a rough estimation.
\begin{figure}
\includegraphics[width=1.0\figurewidth]{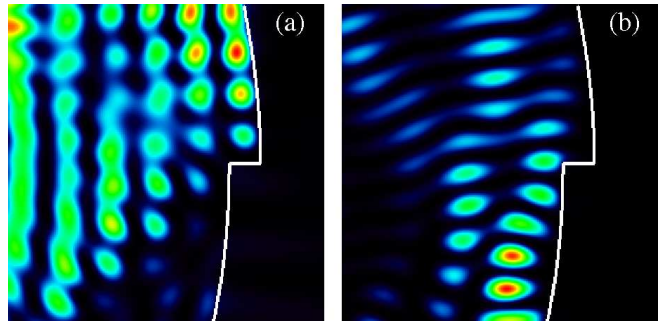}
\caption{(Color online) Magnification of the intensity pattern $|\psi|^2$ near the notch. (a) Quasi-scar mode 2, as in the right panel of Fig~\ref{fig:mode1}. (b) WG-like mode 2, cf. the right panel of Fig~\ref{fig:mode2}.} \label{fig:magnification}
\end{figure}

\section{Dependence on the notch width}
In this section we discuss the dependence on the (relative) notch width
$\varepsilon$. Let us first consider the trivial case $\varepsilon \to 0$, i.e. the circular cavity. Here, pairs of modes are exactly two-fold degenerate. As a consequence the small decay rates $-\imagc{\Omega_j}$ are equal and the level splitting $\Delta\Omega = |\realc{\Omega_1}-\realc{\Omega_2}|$ vanishes. The two modes with angular dependencies $\sim\sin{(m\phi)}$ and $\sim\cos{(m\phi)}$ are obviously orthogonal, i.e. the overlap integral $S$ is zero.

Figure~\ref{fig:dyn} shows the dependence of the decay rates, the level
splitting and the overlap integral as function of~$\varepsilon$. As
expected, the decay rates and the level splitting  increase with increasing
notch width $\varepsilon$. The overlap integral is close to unity, i.e.
$1-S$ is small, in a broad parameter range. This robust behavior is an
interesting finding since usually two parameters have to be carefully adjusted
to an EP in order to get a significant effect. The degree of non-orthogonality 
increases as the scattering becomes more and more asymmetric with increasing
notch width. 
At first glance, the fact that both non-orthogonality and level spacing
increase with increasing $\varepsilon$ seems to be in contradiction with our 
claim that the non-orthogonality is related to an EP where the level splitting
is expected to go to zero. A closer inspection of
Eqs.~(\ref{eq:eigenvalue})-(\ref{eq:eigenvector}), however, shows that there
is no contradiction. According to Eq.~(\ref{eq:eigenvector}) the EP is
approached as $\eta(\varepsilon)$ is decreased. The level splitting, however,
can increase as long as $|V(\varepsilon)|$ increases faster than
$1/\sqrt{\eta(\varepsilon)}$. Roughly speaking, the strength of the scattering
increases faster with the notch width than the asymmetry of the scattering. 
\begin{figure}
\includegraphics[width=1.0\figurewidth]{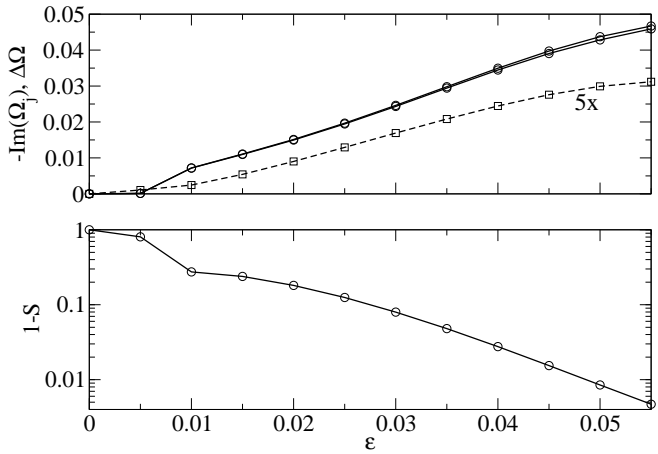}
\caption{The upper panel shows the individual decay rates $-\imagc{\Omega_j}$
  (circles) and the level splitting $\Delta\Omega =
  |\realc{\Omega_1}-\realc{\Omega_2}|$ (scaled by a factor of 5, squares)
  vs. notch width $\varepsilon$.   The corresponding overlap integral $S$
  computed from Eq.~(\ref{eq:overlap}) is plotted in the lower panel. The lines are a guide to the eyes. The pair of modes for the particular case of
$\varepsilon = 0.04$ is the pair of quasi-scars depicted in Fig.~\ref{fig:mode1}.} \label{fig:dyn}
\end{figure}

Figure~\ref{fig:dyn} clearly reveals the direct link between the
almost-degenerate pairs in the spiral for finite $\varepsilon$ and the exact
degenerate pairs of modes with angular momentum number $m\neq 0$ in the circular
cavity. This is another indication that the appearance of non-orthogonal
pairs is a typical feature in the spiral cavity. Only very short-lived modes
which belong to $m=0$-modes of the circle may not follow this scenario.

\section{Summary}
We have demonstrated the appearance of highly non-orthogonal pairs of modes in
optical microspiral cavities. With the help of a non-Hermitian Hamiltonian we
have related this remarkable effect to exceptional points in parameter space.
While usually two parameters have to be adjusted to come close to such a point,
a broad interval of the natural shape parameter of the spiral geometry exhibits
significant non-orthogonality. The physical mechanism behind this effect is the
asymmetric scattering between clockwise and counterclockwise propagating waves.
Simulations of ray dynamics have demonstrated a clear and unexpected ray-wave
correspondence.

We believe that our work is of general importance for the understanding
of open quantum and wave systems, in particular for microcavity lasers.

\section{Acknowledgment}
We would like to thank T.Y.~Kwon for discussions.
Financial support by the DFG research group 760 ``Scattering Systems with Complex Dynamics'' and the DFG Emmy Noether Program is acknowledged.
S.W.~Kim was supported for two years by Pusan National University Research Grant.

\end{document}